# ON RESONANT WAVES IN LATTICES

G. Osharovich[a], M. Ayzenberg-Stepanenko[b]
[a]Bar-Ilan University, Ramat-Gan, Israel
[b]Ben-Gurion University of the Negev, Beer-Sheva, Israel

**Abstract.** Mathematical modeling of resonant waves propagating in 2D periodic infinite lattices is conducted. Rectangular-cell, triangular-cell and hexagonal-cell lattices are considered. Eigenvalues (here eigenfrequencies) of steady-state problems are determined, and dispersion properties of free waves are described. We show that the frequency spectra of the models possess several resonant points located both at the boundary of pass/stop bands and in the interior of a pass band. Boundary value problems with a local monochromatic source are explored, and peculiarities of resonant waves are revealed. Asymptotic solutions are compared with the results of computer simulation. Special attention is given to line-localized primitive waveforms at the resonance frequencies and to the wave beaming phenomena at a resonant excitation.

## 1 INTRODUCTION

Nowadays, nanotechnologies are rapidly developing, and it has become necessary to obtain predictable properties of nanoparticles, nanofibers and nanotubes using various mechanical actions exerted on starting systems. Among various physical characteristics of nanostructures, their waveguide properties are becoming increasingly important allowing theoretical prediction of dynamic stress state and fracture propagation [16].

Mathematical models of nanostructure dynamics are, as a rule, based on periodic lattices of diverse structures and functional impacts. A well-studied wave phenomenon inherent to periodic lattices is that free wave propagation takes place only within certain discrete bands of frequencies known also as pass-bands alternated with stop-bands, where the steady-state wave propagation is forbidden. In the field of *waves in lattices* or, more generally, *waves in periodic media*, the monograph by Brillouin [3] has been basic for subsequent investigations of various theoretical and engineering aspects [15].

During recent decades, this topic got a second wind, when so-called "artificial crystals" used as band-gap materials were discovered. Artificial phononic (or zonic) crystals are periodic lattices or composite structures designed to control sound and vibration waves. Some important results related to band-gaps in phononic crystals can be found, e.g., in [7-8].

In the frequency spectrum of band-gap materials, there exist resonant frequencies, which usually demarcate pass- and stop-bands. In the 1D case, the group velocity of the wave equals zero at these frequencies: there is no steady-state solution corresponding to an external non-self-equilibrated excitation, and the wave energy flows from a source decelerating with time, like heat, and not as a wave [13]. It was shown in [2] that in 2D/3D cases, resonant frequencies also exist *in the interior* of pass bands. Such frequencies differ from those in the 1D case, since the group velocity is zero only for

*some special wave orientations*. Corresponding resonance processes excited in infinite rectangular-cell and triangular-cell lattices by harmonic sources possessing these frequencies were studied in [10-11].

This paper is aimed at obtaining a comparative description of the frequency spectrum of lattices of different structures, proving the existence of so-called Localized Primitive Waveforms (LPWs) in these lattices and describing the peculiarities of resonant waves excited by a local monochromatic source. The LPWs, initially discovered in [2] for square-cell lattices, are "self-equilibrated" standing waves strictly localized on a line of a certain orientation in the lattice, which do not appear along other orientations. The LPWs extracted from the steady-state problem predict a pronounced beam effect obtained from the unsteady-state problem at the resonant excitation in a square-cell lattice which was analytically described [2, 11, 12].

Below we explore free- and time-dependent elastic wave propagation processes in rectangular-cell and hexagonal-cell lattices. The dispersion properties of lattice structures are analytically studied, and computer simulations are conducted in order to reveal the development of the unsteady state pattern. To compare waveguide properties in these two cases, we use known results for rectangular-cell lattices obtained in [12].

The presented numerical solutions (which are of independent significance) are complementary to analytical results and, together with the latter, provide a fairly complete picture of nanostructures dynamics to be analyzed. Note that results of purely computer simulations of wave processes in lattices obtained on the basis of the classical molecular dynamic models and corresponding advanced methods and tools can be found, for example, in [1, 4-6], but an analysis of specific effects for frequency regimes resulting in resonant waves has not been performed.

To explain the term 'resonant wave', we compare frequencies in both 1D finite and infinite simplest uniform mass-spring lattices (MSLs): point particles of mass $M$ are linked by massless elastic bonds of unit length with the stiffness $g$. In Fig. 1, (*a*) and (*b*), such systems are schematically depicted.

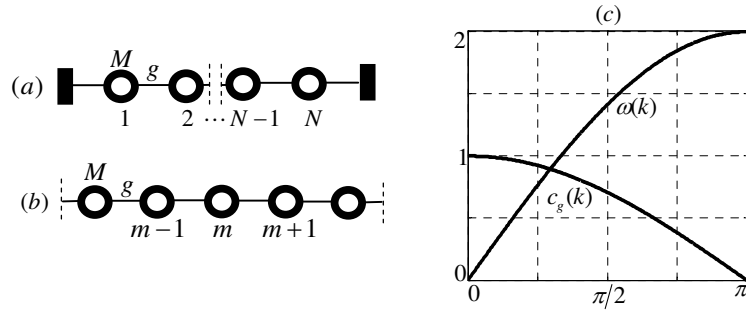

Fig. 1 Finite (*a*) and infinite (*b*) MSLs and dispersion spectra of infinite MSL (*c*)

In the case of a finite system of $N$ oscillators, we have a finite spectrum of $N$ eigenvalues (eigenfrequencies): $\omega_1 < \omega_m < \omega_N$, $(m = 2,\ldots,N-1)$. For physical reasons, the minimal frequency is $\omega_{\min} = \sqrt{2g/NM}$ – all particles perform in-phase motion, so only two boundary springs are subjected to deformation. We can also say that $\lim_{N\to\infty} \omega_N = \sqrt{4g/M}$, while all particles tend to oscillate in anti-phase motion.

Note, finite MSCs are usually used for description of the behavior of locally interconnected dynamical cellular neural networks (see e.g. [9]).

For an infinite MSC we have a continuous frequency spectrum, which can be found using the Floquet approach. The system of homogeneous equations of free waves propagating in the infinite MSL is:

(1) $$\ddot{u}_m - c_0^2 L_2(u_m) = 0, \quad c_0 = \sqrt{g/M}, \quad L_2(u_m) = u_{m+1} - 2u_m + u_{m-1}$$

where $\dot{z} = \partial z / \partial t$, $c_0$ is the sound velocity in an effective homogeneous spring, and $L_2(u_m)$ is the difference operator of the second order.

We seek a solution of the homogeneous system (1) in the form of a traveling wave:

(2) $$u_m(t) = U e^{i(\omega t \pm km)}$$

where $\omega$, $k$ and $\lambda = 2\pi/k$ are, respectively, the temporal frequency, the space frequency and the space wavelength.

Substituting (2) into (1), we obtain the frequency and the group velocity spectra

(3) $$\omega = \pm 2c_0 \sin(k/2), \quad c_g = d\omega/dk = \pm c_0 \cos(k/2).$$

This result has a simple physical sense: the longer the waves ($k$ decreases), the lesser the influence of the waveguide discreteness on wave propagation: $c_g \to \pm c_0$ $(k \to 0)$. The frequency $\omega = 2$ at $k = \pi$ (the edge of the Brillouin zone) demarcates pass and stop bands, while the free waves with frequencies $\omega > 2$ do not propagate. Below, due to symmetry, only the interval $[0, \pi]$ is considered.

In Fig. 1(c), dependencies $\omega(k)$ and $c_g(k)$ are depicted.

Spatial forms of the MSL motion depending on the wave length are determined by the ratio of displacements of neighboring particles. One can see from (2) that in the case of the minimal wavelength, $\lambda = 2 (k = \pi)$, this ratio equals –1: the shortest waves perform anti-phase oscillations.

Differences in resonant patterns for finite and infinite 1D mass-spring lattices are distinct. In the *N*-DOF MSL each $n^{th}$ eigenfrequency is the resonant one. An external excitation of the system with such a frequency results in a linear growth of the solution with time.

In the case of an infinite MSL, we have obtained (see [11]) the following asymptotic solution for problem (1) with zero initial conditions and the action of a local sine source of resonant frequency $u_0(t) = \sin \omega_r t$ ($\omega_r = 2$):

(4) $$u_m(t) \sim \sqrt{t} \left[ F_2(\lambda) \sin(2t - \pi m) - F_1(\lambda) \cos(2t - \pi m) \right], \quad \lambda = 2|m|/\sqrt{t} \quad (t \to \infty, \ |m| \ll t)$$

where the well-known functions in the theory of unsteady waves, $F_1(\lambda)$ and $F_2(\lambda)$ (see, e.g., [11, 13, 14]), are oscillating and spreading with the growth of $\lambda$. The resonant wave (4) propagates along the lattice with the velocity $\sim t^{-1/2}$ that corresponds to the heat propagation law. Within this process, amplitudes $u_m(t)$ increase with time as $t^{1/2}$ contrary to a finite MSL where the linear growth is detected.

In addition to the analytical solution, we present an example of a resonant wave numerically calculated in the infinite MSL subjected to a monochromatic force applied in the zero cross-section of the waveguide: $Q = H(t)\delta(x)\sin\omega_r t$, where $\omega_r = 2$, while $H(t)$ and $\delta(x)$ are Heaviside and Dirac functions. We show how resonant waves are developing under such excitation and whether there is a correspondence between numerical and analytical solutions. Recall that this frequency determines zero group velocity (i.e. zero energy flux) and, for this reason, the steady-state solution is absent. Some results of computer simulations are presented in Fig. 2 [11]. Envelopes of $u_m(t)$ for a set of nodes can be seen in Fig. 2 (a). Dependence $u_0(t)$ is accurately described by asymptotic solution (4) starting virtually from the very beginning. The larger the distance from the considered node to the loading point, the longer the time period needed to reach a good coincidence of the numerical solution with the soluton (4).

Distributions $u_m(t)$ along $n$-axis at $t = 300, 600$, and $900$ are presented in Fig. 2 (b). In accordance with (4), the main perturbations move at the velocity proportional to $t^{-1/2}$, while their amplitudes increase with time as $t^{1/2}$. Calculations show that beginning with $t \approx 200$, solution (4) can be approximated by numerical results with a good accuracy.

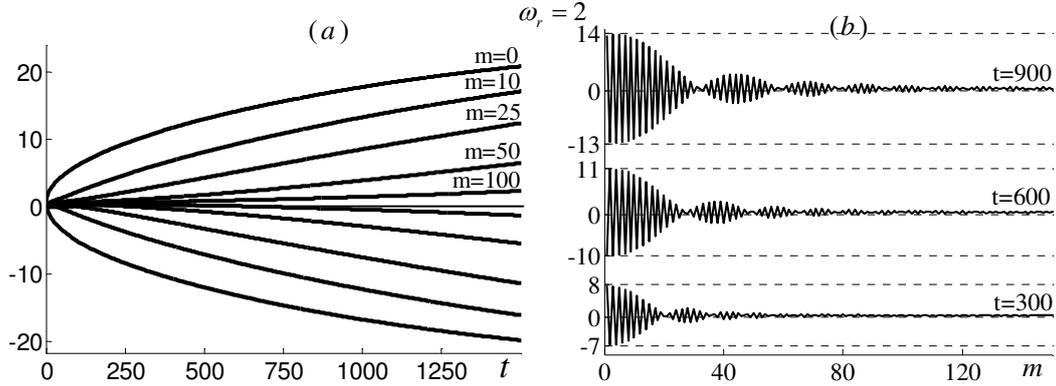

Fig. 2 Resonant process in a MSL: (a) envelopes of displacement oscillations vs. time in nodes $m = 0, 10, 25, 50, 100$; (b) distributions along $m$-axis at $t=300, 600$ and $900$

In 2D cases, in addition to resonances excited by the frequency demarcating pass/stop bands, there exist resonances excited by frequencies located in the interior of a pass band. Below we determine such frequencies and corresponding waveforms and calculate properties of resonant waves in rectangular, hexagonal and triangular lattices.

## 2 RECTANGULAR-CELL LATTICE

In this section we analyze dispersion patterns and anti-plane wave localization processes in a uniform Rectangular-Cell Lattice (RCL). Emphasis is put on the dispersion analysis of wave propagation processes emerging in the lattices under the action of a point harmonic excitation with a resonant frequency.

## 2.1 Dispersion Pattern

Material particles of a lattice are located in nodes $(m,n)$, $m,n = 0, \pm 1, \pm 2, \ldots$, linked by elastic massless bonds – see Fig. 3. We use discrete coordinates $m$ and $n$ together with continuous coordinates $x$ and $y$. The following nomenclature is used: $M$ is the particle mass, $g_x$ and $g_y$ are out-plane normalized stiffnesses of $x$- and $y$-bond, respectively, $l_x$ and $l_y$ are their lengths. We explore transversal oscillations of the lattice (see [12]). First, to explore the dispersion pattern, we are going to obtain a dispersion equation of $\varphi(\omega, k_x, k_y) = 0$ type, where $\omega$ is frequency, $k_x$ and $k_y$ are wave numbers or projections of the wave vector $\mathbf{k}(k_x, k_y)$. Then we analyze dispersion surfaces $\omega = \omega(k_x, k_y)$ and calculate group velocity vectors $\mathbf{c}_g(k_x, k_y)$, $[(c_g)_x, (c_g)_y] = (\omega'_{k_x}, \omega'_{k_y})$.

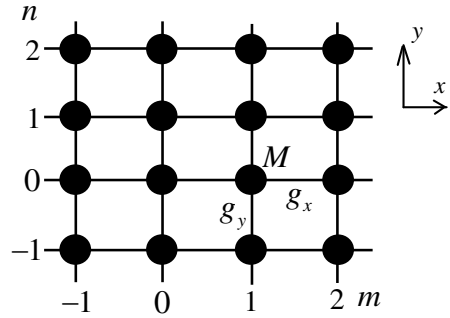

Fig. 3 Rectangular-cell lattice

In a linear approximation, systems of homogeneous equations of the RSL dynamics can be written as follows:

(5) $$M \ddot{u}_{m,n} = g_x \left( u_{m,n+1} + u_{m-1,n} - 2u_{m,n} \right) + g_y \left( u_{m+1,n} + u_{m,n-1} - 2u_{m,n} \right)$$

where $u_{m,n}$ is the out-plane displacement of the $(m,n)$-particle.

We represent a general solution of the system (5) by a superposition of sinusoidal waves

(6) $$u_{m,n}(t) = U_{m,n} e^{i\omega t}, \quad U_{m,n} = U e^{i(k_x x + k_y y)} = U e^{i(k_x m + k_y l \cdot n)} \quad \left( |k_x| \leq \pi, \; |k_y l| \leq \pi \right),$$

where $U$ is constant. Substituting (6) into (5), we obtain the dispersion surface

(7) $$\omega = \sqrt{2\left[ g_x(1 - \cos k_x) + g_y(1 - \cos k_y l) \right] / M}$$

If $x$- and $y$-bonds differ only in their lengths ($l_x \neq l_y$), then, putting $l \equiv l_y$ and taking $l_x$ and $M$ as measurement units, we obtain a dispersion surface for a RCL with a single free parameter $l$,

(8) $$\omega = \sqrt{2\left[ 1 - \cos k_x + (1 - \cos k_y l)/l \right] / M}, \quad \left( |k_x| \leq \pi, \; |k_y l| \leq \pi \right)$$

This surface has a pass-band $\omega \in [0, \omega_r)$ and stop-band $\omega > \omega_r$, where $\omega_r = \sqrt{2(2+2/l)/M}$ is the resonance frequency at the bands interface.

In the case of a lattice with the same bonds ($l = 1$), Eqn. (8) becomes a well-known dispersion relation for a square-cell lattice (SCL),

$$(9) \qquad \omega = \sqrt{2(1-\cos k_x) + 2(1-\cos k_y)} \quad (|k_x| \leq \pi, \; |k_y| \leq \pi),$$

with $\omega_r = \sqrt{8}$ at the pass/stop-bands interface.

Together with the RCL, we introduce its special kind adopted as a Simplified Rectangular-Cell Lattice (SRCL), in which $g_x = g_y$, $l \neq 1$. For a SRCL, the dispersion equation (8) acquires the following form:

$$(10) \qquad \omega = \sqrt{2(1-\cos k_x) + 2(1-\cos k_y l)}, \quad (|k_x| \leq \pi, \; |k_y l| \leq \pi).$$

Then we have revealed similarities and differences inherent to dispersion properties of the considered SCLs and RCLs. For a sinusoidal wave, the group velocity vector $c_g$ (as well as the energy flux) is oriented along an external normal to the equifrequency contour $\omega = \text{const}$. As shown in [2], the contour $\omega = 2$ is resonant for SCLs. Below we show that this contour is also resonant for SRCL.

Fig. 4 depicts dispersion surfaces (*a*) and resonance contours at $\omega = 2$ (*b*) obtained from (6). The contour is rhombic: $k_x \pm l k_y = \pm \pi$, in contrast to the square one, $k_x \pm k_y = \pm \pi$, in the SCL [2],. From (10) we have obtained *x*- and *y*-projections of the group velocity $c_g$ at $\omega = 2$:

$$(11) \qquad (c_g)_x = \frac{\sin k_x}{2}, \quad (c_g)_y = \frac{l \sin k_y l}{2},$$

Here the energy flux along the axes *x* and *y* is absent: $(c_g)_x = 0$ $(k_y = 0)$ and $(c_g)_y = 0$ $(k_x = 0)$. As we show below, the group velocity orientation in the $k_x, k_y$-plane coincides with the orientation of Localized Primitive Waveforms (LPWs) originally discovered in [2]. It follows from (11) that for $\omega = 2$ the group wave velocity value $|c_g|$ and its orientation $\beta$ are

$$(12) \quad |c_g| = \frac{\sqrt{1+l^2}}{2} \sin \frac{\pi}{1+|\tan \alpha|} \quad \left( \alpha = \arctan \frac{l k_y}{k_x} \right), \quad \beta = \arctan \frac{(c_g)_y}{(c_g)_x} = \pm \arctan l,$$

where $\alpha$ and $\beta$ are the phase and group velocity orientations, respectively.

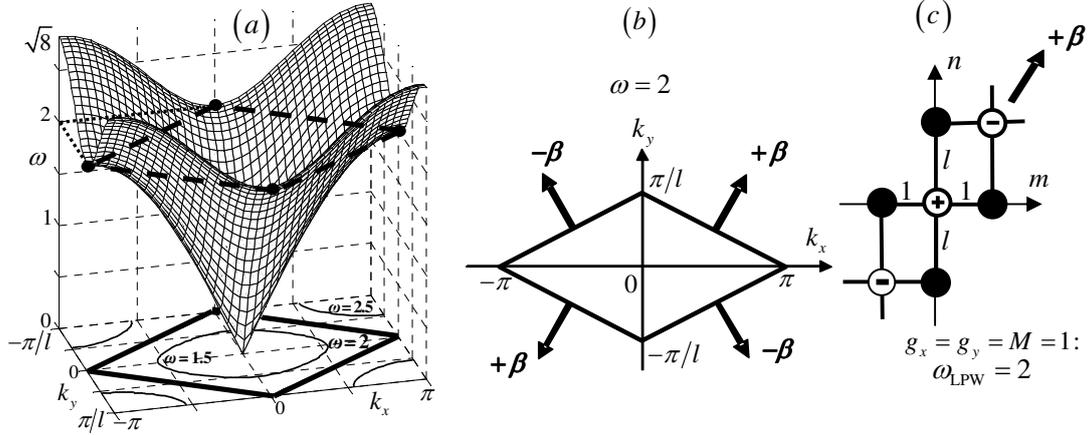

Fig. 4 Dispersion pattern in SRCL: (*a*) dispersion surface (6), (*b*) equifrequency contour for $\omega = 2$ in plane $k_x$, $k_y$ (arrows show the energy flux orientation), (*c*) three-particle-width band. Particles involved in anti-phase oscillations are marked by open circles with '+' or '−', and immobile ones – by black circles

If $l = 1$, we obtain results corresponding to a SCL [2]:

$$(13) \quad |c_g| = \frac{\sqrt{2}}{2} \sin \frac{\pi}{1 + |\tan \alpha|} \quad \left( \alpha = \arctan \frac{k_y}{k_x} \right), \quad \beta = \arctan \left( \frac{\partial \omega}{\partial k_y} \bigg/ \frac{\partial \omega}{\partial k_x} \right) = \pm \frac{\pi}{4}.$$

As one can see from (12), the group velocity is zero only in the following four directions, $\alpha = \pm 0$ and $\alpha = \pm \pi/2$. Directions determining LPW (diagonal) orientations are associated with the angles $\pm \beta$ and shown in Fig. 4 (*b*) by arrows. In Fig. 4 (*c*), for one of the above-mentioned angles, $+\beta$, a three-particle-width band is shown, in which we consider the diagonal $(m,m)$ and a neighboring particle (black circle) connected with two diagonal particles involved in anti-phase oscillations. Their actions on a near-diagonal particle are self-equilibrated, and thus, black particles can be at rest. So the existence of the LPW is a consequence of certain symmetry of the lattice structure. One can see that $\omega = \sqrt{2(g_x + g_y)/M} = 2$ is the LPW frequency as in a SCL (recall that lengths of *x*- and *y*-bonds in a SRCL are different, but their stiffnesses are equal, $g_x = g_y$, and measurement units are $g_x = g_y = M = 1$).

Consider now a RCL possessing bonds of different lengths. The dispersion relation for such a lattice is expressed by (8). Note that LPWs are absent here, but, as shown below, the dispersion pattern has some common points with the above-considered one in SCL (and SRCL).

Projections of the group velocity vector, its module and energy flux directions $\beta$ obtained from (8) are

(14)
$$\left(c_g\right)_x = \frac{\sin k_x}{\omega}, \quad \left(c_g\right)_y = \frac{\sin(lk_y)}{\omega},$$
$$\left|c_g\right| = \frac{1}{\omega}\sqrt{\sin^2 k_x + \sin^2(lk_x)}, \quad \beta = \arctan\left(\frac{\sin(lk_y)}{\sin k_x}\right).$$

Below we also use the expression obtained from (8) for $\beta$ in terms of $\omega$ and $k_x$:

(15)
$$\beta = \arctan\frac{\sqrt{1-\left[1-l\left(\omega^2/2+\cos(k_x)-1\right)\right]^2}}{l\sin(k_x)}.$$

There exist four specific angular points in the dispersion surface ($k_x = \pm\pi, k_y = 0$), where $\omega = 2$ and ($k_x = 0$, $k_y = \pm\pi/l$) – $\omega = 2/\sqrt{l}$, in which the group velocity is equal to zero and, similarly to SCL (SRCL) cases, the directions coinciding with the axes $x$ and $y$ are forbidden for the energy flux.

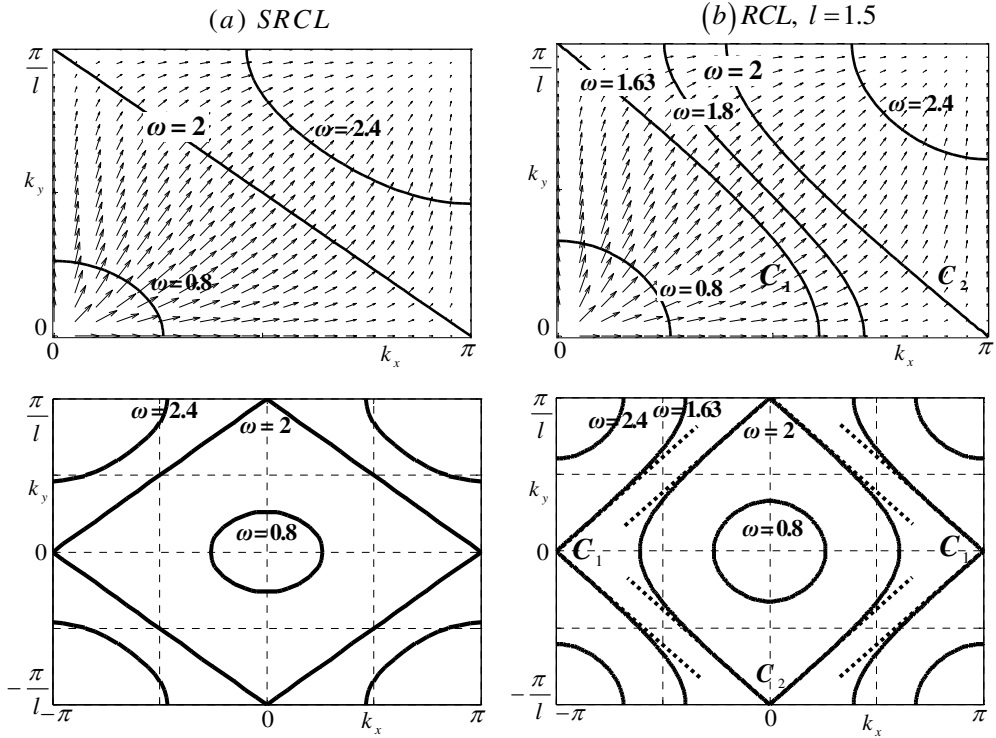

Fig. 5 Group velocity pattern (upper figures) and equifrequency contours (lower figures) in square-cell (*a*) and rectangular-cell (*b*) lattices. Dotted straight lines in the lower picture of (*b*) are tangents to contours at angular points

In two upper pictures in Fig. 5, the first quarter of the plane $k_x, k_y$ is shown with the value and direction of the group velocity expressed by the length and direction of the respective arrow. Besides, some other equifrequency contours are depicted. These contours in the entire Brillouin zone are presented in the lower row of Fig. 5. We compare results for SRL, column (*a*), and RCL $(l = 1.5)$, column (*b*), in order to reveal their similarities and differences. Recall that in a SRCL, resonance frequency $\omega = 2$ determines a rhombic contour $k_x \pm lk_y = \pm \pi$. In the RCL case, equifrequency contours, $C_1$ and $C_2$, are curvilinear, corresponding to the frequencies $\omega_1 = 2/\sqrt{1.5} \approx 1.63$ and $\omega_2 = 2$, respectively.

The analysis of relations (14) and (15) shows that for relatively low frequencies, there are no preferable directions of $c_g$. If the frequency increases, the dependence of $|c_g|$ and $\beta$ on coordinates $k_x$ and $k_y$ becomes sensible.

Consider the SRL case, Fig. 5(*a*). If the frequency tends to a resonant one, $\omega = 2$, values of $|c_g|$ tend to zero in the vicinity of the above-mentioned angular points of the contour, $k_x = \pm \pi, k_y = 0$ and $k_x = 0, k_y = \pm \pi/l$. When $\omega_0 = 2$, the directions of $c_g$ with wave numbers $k_x \sim 0$ ($k_y \sim 0$) turn to the right (left) at an angle $\pi/2$. Such behavior of the free wave pattern is related to the formation of *caustics*. In the RCL case, the caustic appears in the same angular points, which are located now in different contours, $C_1$ and $C_2$.

As shown in [12], sources with the frequencies $\omega_0 = \omega_1$ and $\omega_0 = \omega_2$ excite resonance phenomena of a pronounced beaming character of the spatial wave pattern. With increasing $\omega_0$, the process of $c_g$ transition consists of two parts:

(*i*) If $\omega_0$ tends to $\omega_1 = 2/\sqrt{l} = 1.63$, the value of $|c_g|$ in the vicinity of $k_x \sim 0$ decreases with increasing $k_y$. It tends to zero if $\omega_0 = \omega_1$ and $k_y \to \pi/l$, and $c_g$ orientation (angle $\beta$) sharply turns from $\beta = \pi/2$ to $\beta = 0$. If $k_x$ increases, the value of $|c_g|$ in the contour $C_1$ also increases, while the change in $\beta$ orientation obeys the requirement for the vector $c_g$ to be normal to $C_1$.

(*ii*) With further increase in $\omega$ while approaching to $\omega_2 = 2$ (contour $C_2$) and passing through it, the group velocity pattern is similar to that described above in case (*i*), but it is realized now near the domain ($\omega = \omega_2$, $k_x \to \pi$), and the vector $c_g$ turns from $\beta = 0$ to $\beta = \pi/2$.

The analysis of the dispersion pattern shows that most parts of the contours $C_1$ and $C_2$ can be approximated by tangents (which are mutually parallel) at angular points. From (15) we have obtained $c_g$ orientation corresponding to these tangents:

(16) $$\beta = \beta_* = \arctan(1/\sqrt{l})$$

This angle is prevalent for wave propagation from a local source with the frequencies $\omega_1$ and $\omega_2$ to the periphery. A certain part of the wave is scattered inside the interval $0 < \beta < \beta_*$, while the interval $\beta_* < \beta \leq \pi/2$ determines forbidden directions for the energy flux.

## 2.2 Unsteady-state dynamics

We have analyzed wave patterns at kinematic excitation, $u_{0,0} = U \sin \omega_0 t$, and/or force excitation, $F_{0,0} = F \sin \omega_0 t$, applied to a particle $m = n = 0$ at $t = 0$. Here $U$ and $F$ are excited amplitudes (below we set $U = 1$ and $F = 1$). In computer simulations, we use Eqn. (5) with zero initial conditions and above-mentioned loadings. An explicit finite-difference scheme has been used in a finite domain of lattice nodes, while the domain boundaries are chosen at such a long distance from the source that their influence on the space region of interest has not been detected. The time step is taken two orders less than the time measurement unit, which (as calculations show) allows the accuracy level $\sim 10^{-4}$.

First we consider a SCL as a partial case of the RCL. At the resonant excitation, $\omega_0 = 2$, a phenomenon of "spatial wave separation" is detected in Fig. 6. As one can see, a sufficiently large part of the source energy is captured by the diagonal line $n = m$ ($\beta = \pi/4$). In accordance with the dispersion analysis and relations (13), $c_g = \sqrt{2}$ for $\omega = 2$. At the same time, oscillations along the directions $n = m/2$ and $n = 0$ are practically locked in the source vicinity.

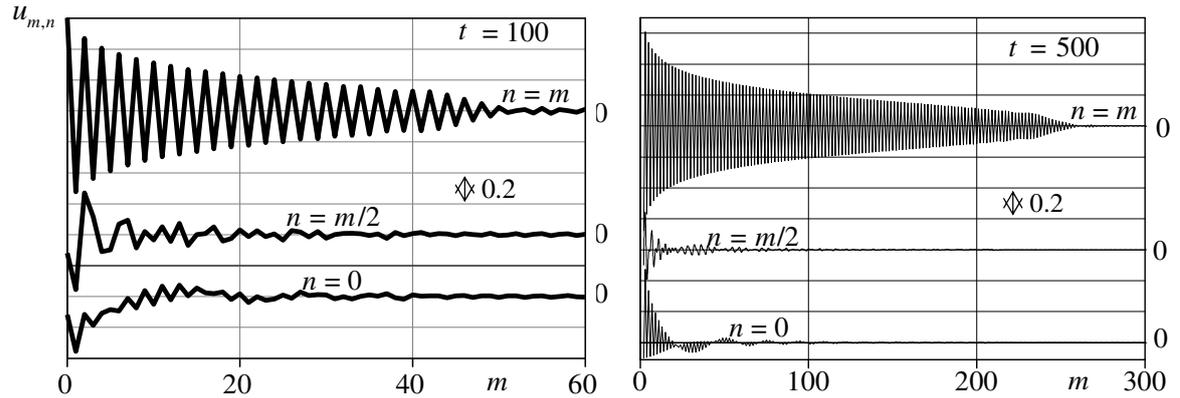

Fig. 6 Distributions of displacements along three directions: $n = m$, $n = m/2$ and $n = 0$ at the force excitation with $\omega_0 = 2$. Pronounced beaming pattern of wave propagation along the direction $n = m$ is detected

Note that there are no noticeable differences in kinematic and force excitations (this conclusion remains valid in the examples of calculation presented below).

Some results of the star-beaming phenomenon are presented in Fig. 7 (*a*) and (*b*) for SCL and SRCL (with $l = 1.5$), respectively. The structures are subjected to a local kinematic excitation with resonance frequency $\omega_0 = 2$. The star-like wave contours are plotted in such a way that the envelopes of oscillations in the outer area, $|U_{m,n}|$, are lower than 10% of the source amplitude – $|U_{m,n}| < 0.1$.

In Fig. 8, we present examples of the same perturbation pattern in a RCL (the parameter of the cell form is $l = 1.5$) at $t = 250$. Kinematic excitation with several values of $\omega_0$ was applied. Here spatial wave patterns are more saturated than in the SCL-SRCL case (compare with Fig. 7). First, in accordance with the orientation (16), a star-like pattern appears for two frequencies $\omega_0 = \omega_1 = 1.63$ and $\omega_0 = \omega_2 = 2$. Calculations show that for these frequencies a resonant character of oscillations is realized, similar to that in SCLs.

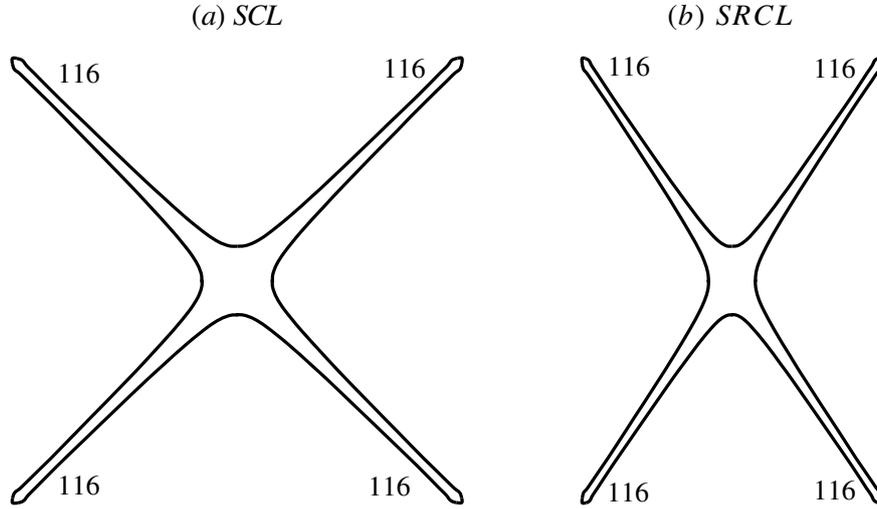

Fig. 7 Star-like beaming pattern at the resonance frequency $\omega_0 = 2$ in square-cell (*a*) and simplified rectangular-cell (*b*) lattices at $t = 250$. Outside the stars, displacements of nodes remain less than 10% of the maximal value in the source

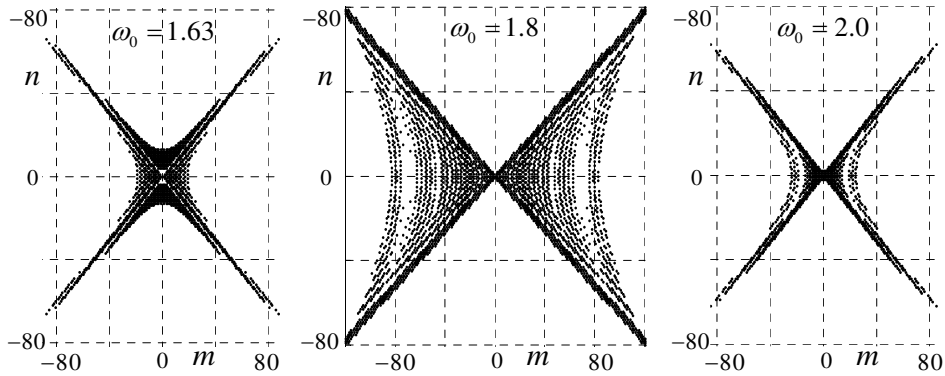

Fig. 8 Beaming wave patterns in the rectangular-cell lattice at $t = 250$. Points represent particles with $|U_{m,n}| \geq 0.1$

The spatial character of wave propagation with frequencies within the interval $(\omega_1, \omega_2)$ is restricted by forbidden directions in compliance with (15) and the above-presented dispersion analysis.

Finally, we have compared the results of computer simulations conducted in the case of the resonance frequency located at the interface of pass- and stop-bands: $\omega_0 = \omega_r = \sqrt{2(2+2/l)}$. In Fig. 9, modules of envelopes, $|U_{m,n}|$, in diverse nodes (*m,n*) are depicted as functions of time for several values of *l* (force source functions). The main difference of the wave response in the considered case from that discussed above, where exciting frequencies are within the pass-band, is that in the former there are no preferable directions for wave propagation. Besides, resonant growth of perturbations with time is relatively slower.

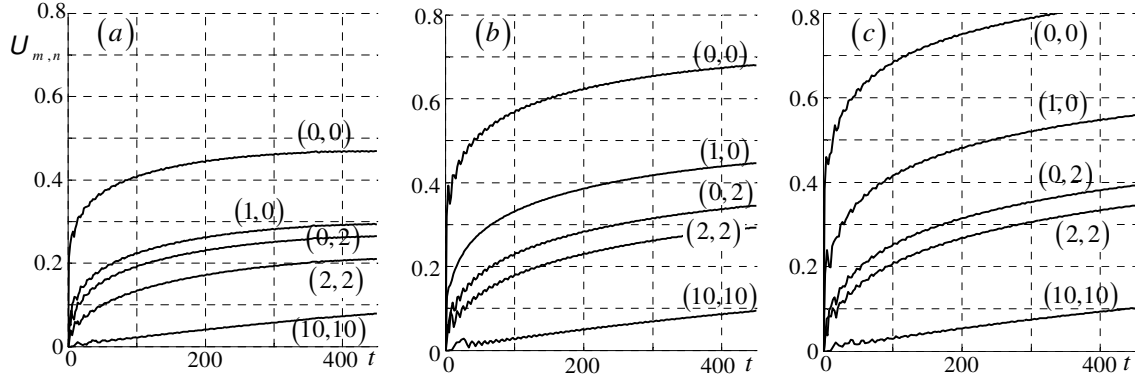

Fig. 9 Envelopes $|U_{m,n}|$ vs. time in RCLs at excitations with frequency $\omega_0$ demarcating pass and stop bands: (*a*) $l = 0.5$, $\omega_0 = \sqrt{12}$; (*b*) $l = 1$, $\omega_0 = \sqrt{8}$; (*c*) $l = 1.5$, $\omega_0 = \sqrt{20/3}$

## 3 HEXAGONAL-CELL LATTICE

### 3.1 Dispersion pattern

Consider a uniform hexagonal-cell lattice HCL of material particles at nodes $(m,n)$, $m, n = 0, \pm 1, \pm 2, \ldots$, linked by elastic massless bonds – see Fig. 10. We use discrete rhombic Cartesian coordinates $m$ and $n$ together with continuous rectangular coordinates $x$ and $y$. As above, we consider out-plane oscillations of the lattice. The generating element of the lattice, rhombic cell $[m,n]$, is bounded by coordinate lines $m$, $n$, $m+1$ and $n+1$. It consists of two nodes, whose displacements are denoted $u$ and $v$. The $v$-nodes are located at the intersection of coordinate lines $m$ and $n$, while $u$-nodes are located leftward of the correspondent $v$-node. Each $v$-node is connected to the three nearest $u$-nodes, and vice versa, each $u$-node is connected to the three nearest $v$-nodes. Such a consideration results in the existence of two oscillating modes.

Let the distance between two neighboring nodes, particle masses and stiffnesses of connecting bonds be measurement units: $M = g = a = 1$. Then, with the cell geometry in mind, transverse motion of particles arranged in a hexagonal lattice is described by the following system:

$$(17) \quad \begin{aligned} \ddot{u}_{m,n} &= v_{m,n} + v_{m,n-1} + v_{m-1,n} - 3u_{m,n} \\ \ddot{v}_{m,n} &= u_{m,n} + u_{m,n+1} + u_{m+1,n} - 3v_{m,n} \end{aligned}.$$

As in [16], we assume harmonic solutions of (17) in the conventional form of a plane wave (in $xy$-plane)

$$(18) \quad (u_{m,n}, v_{m,n}) = (U, V) \cdot \exp\left(i\left(\omega t + n\vec{k}\cdot\vec{a}_1 + m\vec{k}\cdot\vec{a}_2\right)\right),$$

where $\vec{a}_1 = \sqrt{3}/2\,\hat{x} + 1/2\,\hat{y}$, $\vec{a}_2 = \sqrt{3}/2\,\hat{x} - 1/2\,\hat{y}$ ($\hat{x}$ and $\hat{y}$ are unit vectors).

After substituting (18) into (17), we obtain a linear system, whose nontrivial solution determines a dispersion equation:

$$\omega_{\text{I,II}} = \sqrt{3 \mp \sqrt{1 + 4\cos(k_y/2)\left(\cos(k_x\sqrt{3}/2) + \cos(k_y/2)\right)}} \qquad (19)$$

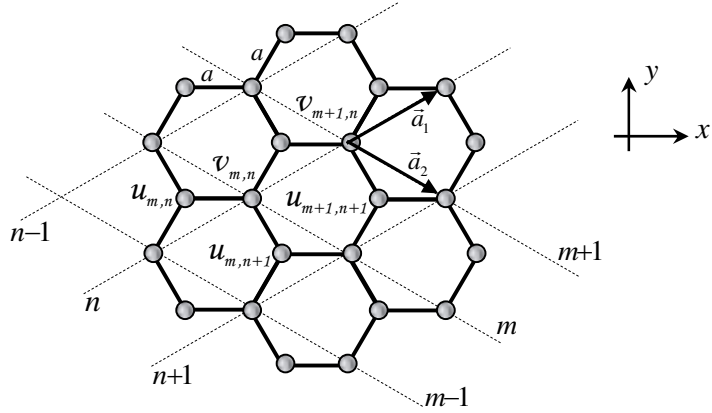

Fig. 10 Hexagonal-cell lattice

where signs '−' and '+' correspond to the first (acoustical) and second (optical) modes.

Eqn. (19) shows that a complete stop-band between modes is absent: their dispersion surfaces are connected in four so-called conical points (CP), which are obtained by equating $\omega_{\text{I}} = \omega_{\text{II}}$. Coordinates of CPs are $[k_x, k_y] = [\pm 2\pi/\sqrt{3}, \pm 2\pi/3]$, and $\omega_{CP} = \sqrt{3}$ is the CP frequency. The point $k_x = k_y = 0$ determines resonant frequency $\omega = \sqrt{6}$ demarcating pass band, $\omega < \sqrt{6}$, and stop band, $\omega > \sqrt{6}$.

Below we show that LPWs are realized at the same single contour $k_y = \pm\pi \cup k_y = \pm 2\pi \pm \sqrt{3}k_x$ corresponding to two resonant frequencies $\omega = \sqrt{2}$ (mode I) and $\omega = 2$ (mode II).

In Fig. 11, dispersion surfaces of modes I and II are depicted together with sets of equifrequency contours for each of the modes.

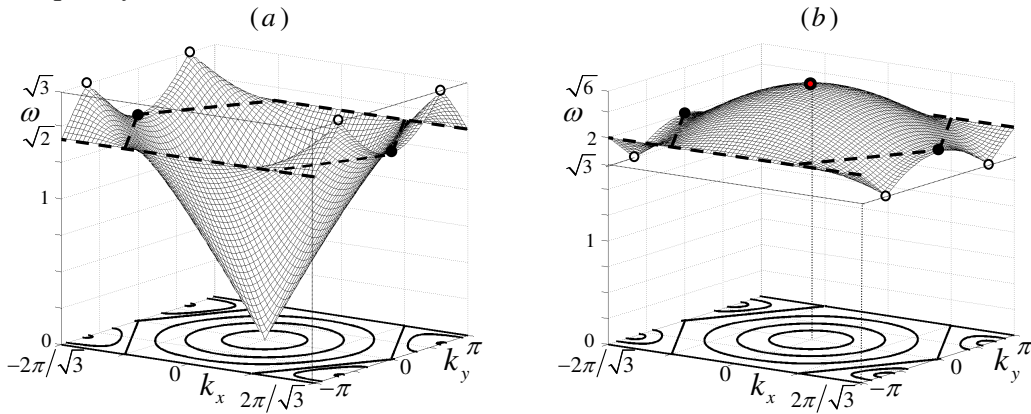

Fig. 11 Dispersion surfaces and equifrequency contours for modes I – (*a*) and II – (*b*). Black circles are angular points of contours, in which group velocities are zero. Open circles are CPs, in which surfaces of modes I and II are connected

To evaluate parameters of the energy flux depending on the frequency and the wave vector, we have obtained from (19) group velocities (absolute values, $x$- and $y$-projections and directions, $\beta$):

$$\left(c_{g,x}\right)_{\mathrm{I,II}} = \frac{\sqrt{3}\cos\left(k_y/2\right)\sin\left(\sqrt{3}k_x/2\right)}{2\omega_{\mathrm{I,II}}\phi\left(k_x,k_y\right)}, \quad \left(c_{g,y}\right)_{\mathrm{I,II}} = \frac{\sin\left(k_y/2\right)\cos\left(\sqrt{3}k_x/2\right)+\sin\left(k_y\right)}{2\omega_{\mathrm{I,II}}\phi\left(k_x,k_y\right)},$$

$$\left(\phi\left(k_x,k_y\right) = \sqrt{1+4\cos\left(k_y/2\right)\left[\cos\left(k_x\sqrt{3}/2\right)+\cos\left(k_y/2\right)\right]}\right),$$

(20) $\quad \left|c_g\right|_{\mathrm{I,II}} = \sqrt{\left(c_{g,x}\right)_{\mathrm{I,II}}^2 + \left(c_{g,x}\right)_{\mathrm{I,II}}^2}, \quad \beta_{\mathrm{I,II}} = \arctan\dfrac{3\cos^2\left(k_y/2\right)\sin^2\left(\sqrt{3}k_x/2\right)}{\sin\left(k_y/2\right)\cos\left(\sqrt{3}k_x/2\right)+\sin\left(k_y\right)}.$

Consider the equifrequency contour $\left[k_y = \pm\pi \cup k_y = \pm 2\pi \pm \sqrt{3}k_x \left(-\pi \leq k_x \leq \pi\right)\right]$ at $\omega = \sqrt{2}$, which is of special interest:

(21)
$$k_y = \pm\pi \quad \Rightarrow \quad \left|c_g\right|_{\mathrm{I,II}} = \frac{\alpha_{\mathrm{I,II}}\left|\cos\left(\sqrt{3}k_x/2\right)\right|}{4}, \quad \beta = \pm\frac{\pi}{2}$$

$$k_y = \pm 2\pi \mp \sqrt{3}k_x \quad \Rightarrow \quad \left|c_g\right|_{\mathrm{I,II}} = \frac{\alpha_{\mathrm{I,II}}\left|\sin\sqrt{3}k_x\right|}{4}, \quad \beta = \pm\frac{\pi}{6}$$

$$\left(\alpha_{\mathrm{I}} = \sqrt{2},\ \alpha_{\mathrm{II}} = 1\right)$$

The group velocity orientation for mode II is opposite to that for mode I while their absolute values differ by a factor of $\sqrt{2}$. As mentioned above, the group velocity orientation in $k_x, k_y$-plane coincides with the orientation of LPWs.

In the upper row of Fig. 12, first quarters of the $xy$-plane are shown; the value and direction of the group velocity are expressed by the length and direction of the respective arrow; also parts of equifrequency contours can be seen. These contours in the entire Brillouin zone are presented in the lower row of Fig. 12.

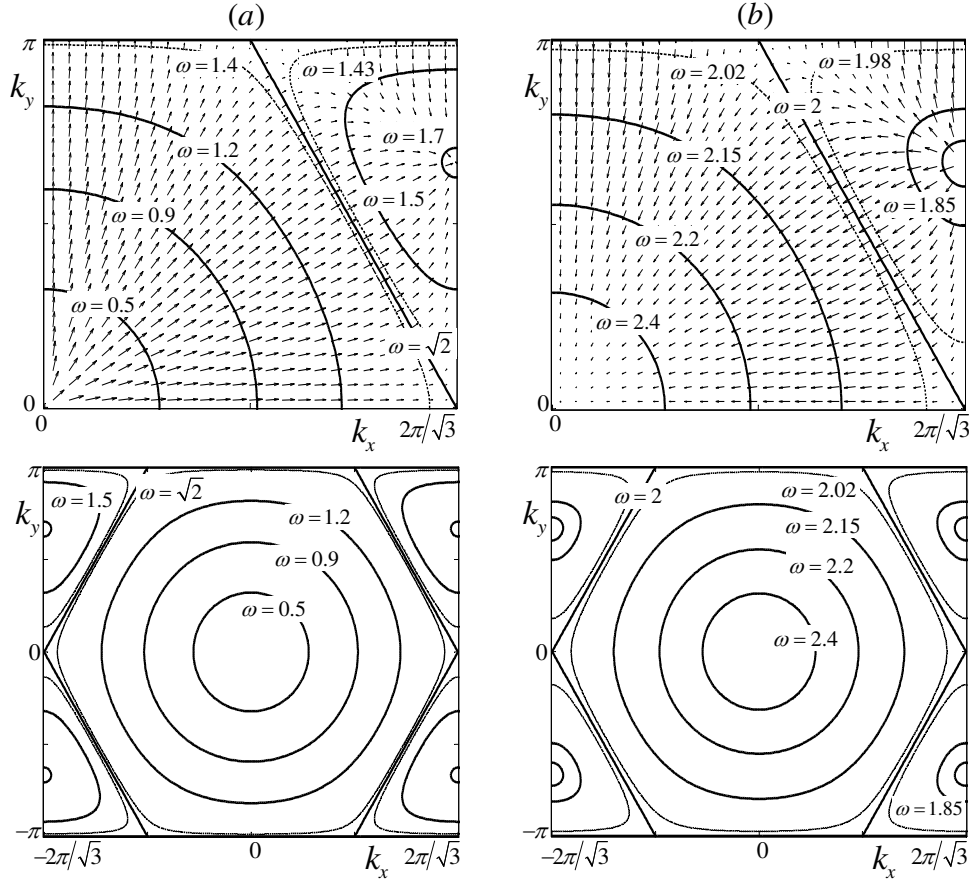

Fig. 12 Dispersion pattern in the HCL. Group velocities (the upper row) in the first quarters of the plane $k_x$, $k_y$ and equifrequency contours in the Brillouin zone (the lower row): ($a$) – mode I, ($b$) – mode II

We compare the results for the acoustic mode, column ($a$), and the optic mode, column ($b$), in order to reveal their differences. In the case of long waves $k_x, k_y \to 0$, the mode I has a maximal value $c_g = 1/2$, while the mode II determines $c_g = 0$; the opposite directions of velocities in these two cases are detected. The group velocity in CP is $c_g = 1/4$ in both modes.

The LPW frequencies in HCLs can be obtained using considerations similar to those used above in the SRCL case. In Fig. 13, two-particle-width bands are depicted allowing LPW frequencies for mode I ($a$) and mode II ($b$) to be evaluated; direction $\beta$ corresponds to one of three directions in which the energy flux possesses the maximal value. First, consider a band shown in Fig. 13($a$). The particles of $u$- and $v$- families of each cell move in-phase (this is denoted by the same signs), according to the oscillation form of mode I. The motion of particles $u$ and $v$ is realized in a neighboring cell also in-phase, but with the opposite sign. The actions of two neighboring pairs on intermediate black particles are self-equilibrated, and thus, the latter can be at rest. The oscillation frequency of the cell is $\omega_I = \sqrt{4g/2M} = \sqrt{2}$ in the considered case (recall that $g$ and $M$ are measurement units).

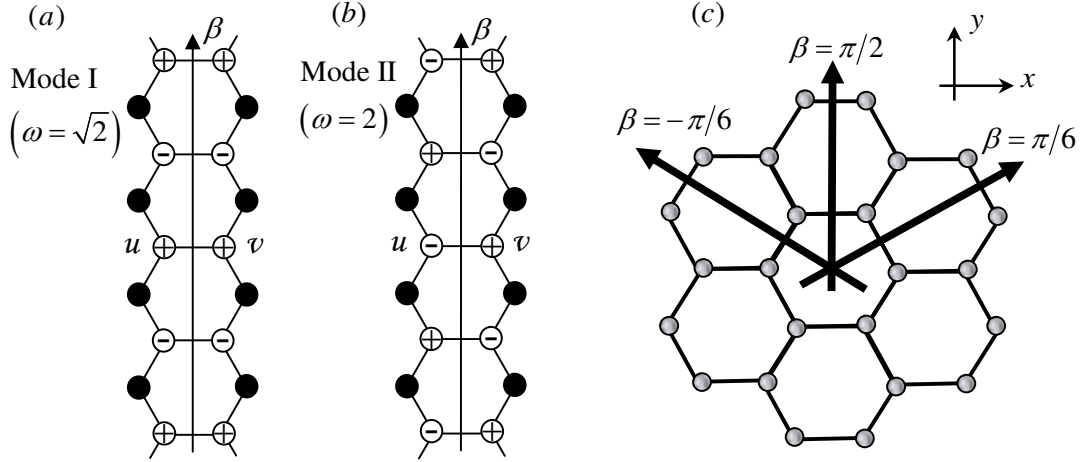

Fig. 13 LPWs of mode I (*a*) and mode II (*b*); (*c*) – energy flux orientations $\beta$

A similar consideration applied to anti-phase oscillations inherent to mode II, Fig. 13 (*b*), allows to obtain the corresponding LPW frequency as $\omega_{II} = \sqrt{4g/M} = 2$: the total stiffness in this case is equal to $4g$ ($2g$ is the sum of stiffnesses of two bonds linking a moving particle to immobile ones, and $2g$ is the stiffness of the half-bonds linking moving particles with each other). The LPW orientations are shown in Fig. 13 (*c*).

In addition to the LPWs, a special oscillation form exists, in which particles of $u(v)$-family are immobile, while neighboring particles of $v(u)$-family oscillate in anti-phase. This form has the CP frequency $\omega = \sqrt{3}$. Finally, note a special form in which neighboring particles oscillate in anti-phase. The frequency of this form turns out be the pass/stop-bands interface $\omega = \sqrt{6}$. The latter case is an analogue of the simplest anti-phase resonant oscillations in a one-dimensional mass-spring chain with the frequency $\omega = 2$ demarcating pass- and stop-bands.

**3.2 Unsteady-state dynamics**
Some results of computer simulations with HCLs are presented below in Fig. 14 and Fig. 15. Calculations were conducted in the case of force excitation, and the location of particle $u_{0,0}$ was associated with the origin of rectangular coordinates $x = y = 0$. Envelopes of $u_{m,n}$ and $v_{m,n}$ depicted in Fig. 14 correspond to two cells: *1* – $(m,n) = (23,0)$ and *2* – $(m,n) = (-15,26)$. They are obtained at the following frequencies $\omega_0$: $\omega_0 = \sqrt{2}$ and $\omega_0 = 2$ (note that those are resonant frequencies which excite LPWs). In rectangular coordinates, cell *1* – $(x,y) \approx (35,20)$ and cell *2* – $(x,y) \approx (16.5,-35.5)$ are located at practically the same distance, $d \approx 40$, from the source, and cell *1* is located on the resonant ray $(\beta = \pi/6)$, while cell *2* is located approximately in the middle between two resonant rays $\beta = \pi/6$ and $\beta = \pi/2$. In the case of $\omega_0 = \sqrt{6}$ (recall that this frequency is located at pass/stop-band intersection), cell 1: $(m,n) = (7,0)$ and cell 2: $(m,n) = (-5,8)$ were chosen. They are located at practically the same distance, $d \approx 12$, from the source.

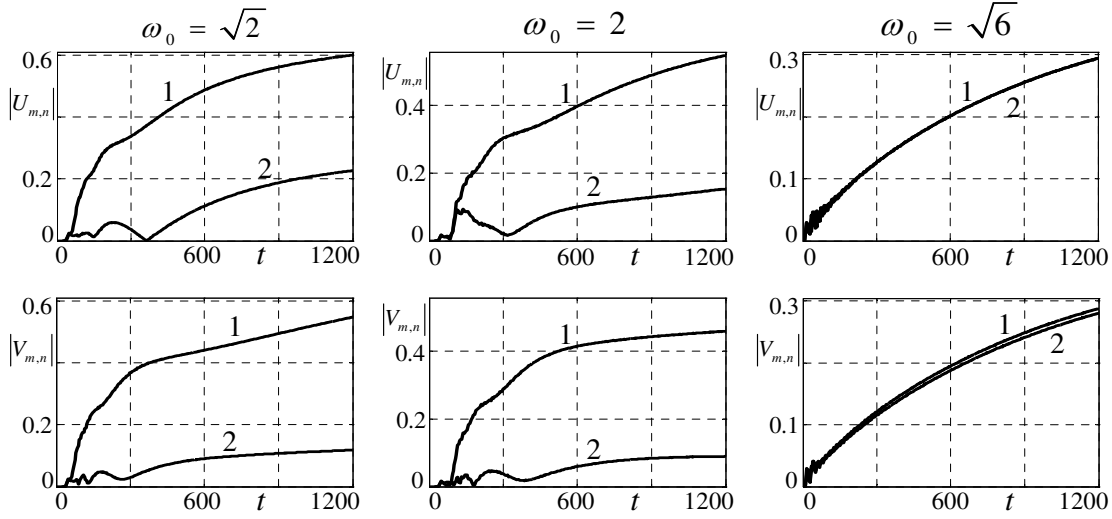

Fig. 14 Envelopes of displacements in diverse nodes at a set of frequencies $\omega_0$

The pattern described above is also realized in symmetric directions (with respect to lattice structure). The presented results, together with the results of additional calculations, allow the formulation of the following conclusions:

1. Relatively long waves have no preferred directions of propagation (up to values close to the first resonance, $\omega_0 = \sqrt{2}$). The influence of lattice structure does not appear, and waves propagate as in a corresponding effective homogeneous solid. Finally, the steady-state solution is reached relatively rapidly.

2. The wave pattern is drastically changed at the resonant frequency $\omega_0 = \sqrt{2}$ associated with the LPW for mode I – a detectable growth of amplitudes in the star rays and a weak response in intermediate regions are detected.

3. If the exciting frequency coincides with $\omega = \sqrt{3}$, a surprising wave pattern is detected at first glance: while oscillations of $u$-particles of the family, to which the excited particle $u_{0,0}$ belongs, tend to zero with time, oscillations of $v$-particle tend to the steady state limit. On the other hand, it can be seen from the system for $U$ and $V$, whose nontrivial solution determines dispersion equation (14), that the ratio $U/V \to 0$ (or $V/U \to 0$) if $\omega \to \sqrt{3}$. So such a pattern obtained in the unsteady state solution could be expected.

4. The LPW resonant frequency for mode II, $\omega_0 = 2$, results in practically the same wave pattern as the one considered above for mode I. The difference is that in accordance with the dispersion pattern presented in (21) the main perturbations arrive at the observation points with a delay (compare with the results for $\omega_0 = \sqrt{2}$).

5. Frequency $\omega_0 = 2.2$ located between two resonances $2 < \omega < \sqrt{6}$ results in the same qualitative process as in the case a low-frequency excitation $\omega_0 = 1.2$: the wave has no preferred directions of propagation, and the steady state regime is reached relatively quickly.

6. Amplitudes of low-frequency long-wavelength resonance excited by the frequency demarcating pass- and stop-bands, $\omega = \sqrt{6}$, increase relatively slowly, and the resonant pattern is detected as soon as the perturbations arrive at the reference point. As in the 1D case (see [11, 13]), the group velocity is equal to zero at this frequency – there is no steady-state solution corresponding to an external non-self-equilibrated excitation, and the wave energy flows from a source decelerating with time, like heat, and not as a wave.

In Fig. 15 star-like beaming patterns of wave propagation process are shown at source frequencies $\omega_0 = \sqrt{2}$ and $\omega_0 = 2$. It is of interest that the beaming is more pronounced in the case of mode II. Such a phenomenon will be the point of further research.

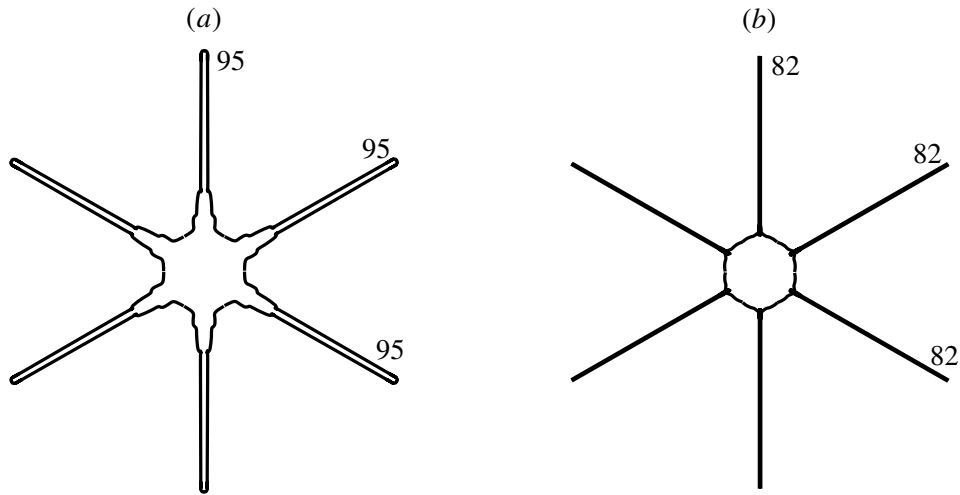

Fig. 15 Star-like beaming pattern in the hexagonal-cell lattice at the resonant frequencies (kinematic excitation): (*a*) mode I, $\omega_0 = \sqrt{2}$ and (*b*) mode II, $\omega_0 = 2$ at $t = 400$. Outside the contours, maximal displacements of nodes remain less than 10% of the maximal value in the source

## 4 TRIANGULAR-CELL LATTICE

Two considered kinds of triangular-cell lattices are shown in Fig. 16 (*a*) and (*b*), respectively, a lattice generated by an equilateral triangle (ETL), and a lattice generated by a right-angled triangle (RTL).

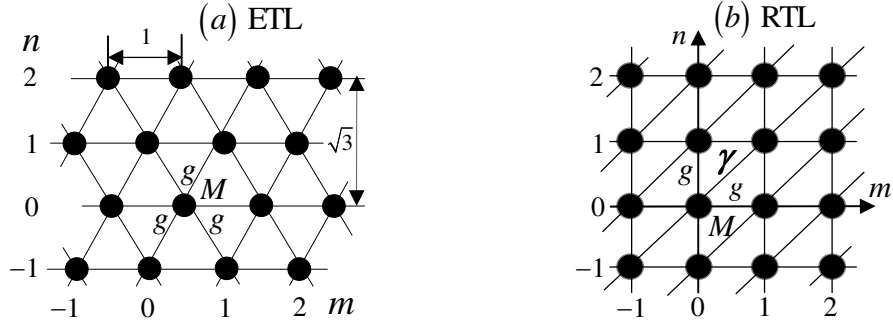

Fig. 16 Triangular-cell lattices: (*a*) ETL – a lattice of equilateral triangles, (*b*) RTL – a lattice of right-angled triangles.

**4.1 Waves in ETLs**

We consider a transverse motion of an ETL consisting of particles (with the mass $M$) located at the points $x = m + n/2$, $y = \sqrt{3}n/2$ $(m, n = 0, \pm1, \pm2, \ldots)$ and connected by elastic massless springs of the stiffness $g$. Let these natural parameters be measurement units: $M = g = 1$. Then a non-dimensional homogeneous system of dynamic equations for such a lattice $(m, n \in Z)$ is

(22)  $$\ddot{u}_{m,n} = u_{m,n+1} + u_{m+1,n} + u_{m-1,n} + u_{m,n-1} + u_{m-1,n+1} + u_{m+1,n-1} - 6u_{m,n}.$$

From (22) we have obtained the dispersion surface

(23)  $$\omega = \sqrt{8 - 4\cos\left(\frac{k_x}{2}\right)\left[\cos\left(\frac{k_x}{2}\right) + \cos\left(\frac{\sqrt{3}k_y}{2}\right)\right]}, \quad -\pi \leq k_x \leq \pi, \quad -\frac{2\pi}{\sqrt{3}} \leq k_y \leq \frac{2\pi}{\sqrt{3}}$$

Here, the (resonance) frequency demarcating pass- and stop-bands is $\omega_r = 3$. Four resonant points of this frequency have *y*-coordinates at the end the Brillouin zone $q_y = \pm 2\pi/\sqrt{3}$, while *x*-coordinates, $q_x = \pm 2\pi/3$ (which is of interest) are found inside it [10]. The LPW frequency located inside the pass-band is $\omega = \sqrt{8}$ [2].

In Fig. 17 (*a*), a resonant equifrequency contour at $\omega = \sqrt{8}$ is plotted along with two others contours corresponding to $\omega = 2.0$ and 2.9. The group velocity obtained from (23) at $\omega = \sqrt{8}$ is

(24)  $$|c_g| = \frac{1}{2\sqrt{8}} \cos\left(\frac{\sqrt{3}\pi}{2}\tan\alpha\right), \quad \beta = 0, \quad -\pi/6 < \alpha = \arctan(k_y/k_x) < \pi/6$$

and the energy flux in the sector $|\alpha| < \pi/6$ is directed along $\pm x$. For the remaining five sectors, $2k - 1 < 6\alpha/\pi < 2k + 1$ $(k = 1, \ldots, 5)$, we obtain the same value for $|c_g|$ for five orientations $\beta_k = k\pi/3$. Thus, just as for rectangular lattices, the group velocity has the orientation of the nearest LPW, and $c_g = 0$ at the vertices of the hexagon, where $c_g$ direction is discontinuously changing.

The existing LPWs, whose orientations are associated with each of the three bond lines, can be detected in a three-particle-width band presented in Fig. 17 (*b*) for $\beta = \pi/3$. Here, as in the case of the SRL, Fig. 4 (*c*), particles inside the band involved in anti-phase oscillations are shown by open circles with signs '+' and '−', while black circles correspond to immobile particles. The action of oscillating particles on a neighboring one is self-equilibrated. The eigenfrequency of an oscillating particle is $\omega = \sqrt{G_\Sigma/M}$, where $M = 1$ and $G_\Sigma$ is the total stiffness of bond system equal to 8 (it consists of four bonds with unit stiffness connecting an oscillating particle with neighboring immobile ones, and two half-bonds with the stiffness equal to 2 each). Thus, the system eigenfrequency is $\omega = \sqrt{8}$.

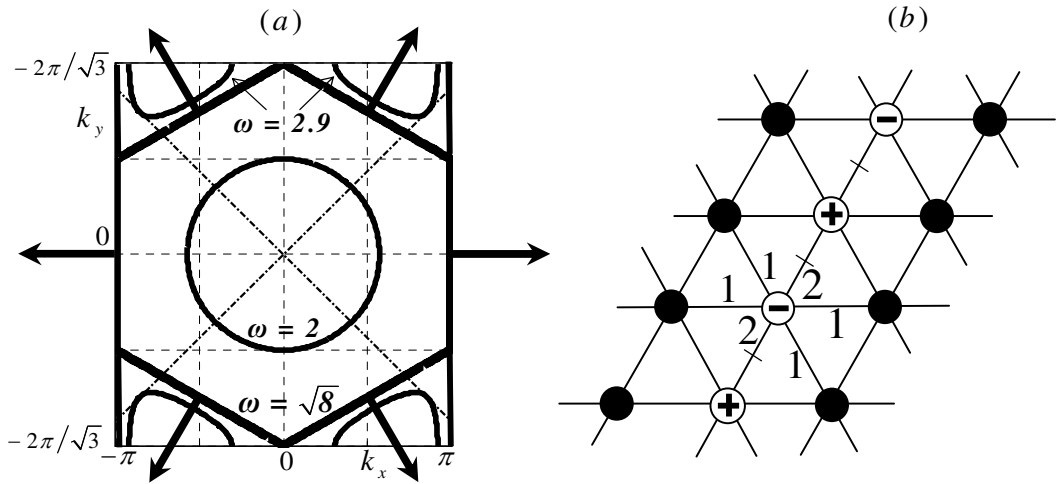

Fig. 17 Dispersion pattern in ETLs: (*a*) equifrequency contours and energy flux orientations (bold arrows) in the resonant case $\omega = \sqrt{8}$, (*b*) LPWs in a three-particle-width band.

### 4.2 Waves in RTLs

Now we examine the transverse motion of a RTL. In this case, the system of homogeneous dynamic equations and the corresponding dispersion relation are as follows (parameters *M* and *g* being taken as measurement units: $M = g = 1$):

$$(25) \quad \ddot{u}_{m,n} = u_{m,n+1} + u_{m+1,n} + u_{m-1,n} + u_{m,n-1} + \gamma\left(u_{m+1,n+1} + u_{m-1,n-1}\right) - 6u_{m,n}$$

$$(26) \quad \omega = \sqrt{4 - 2\cos k_x - 2\cos k_y + 2\gamma\left[1 - \cos(k_x + k_y)\right]}$$

In contrast to the square-cell lattices and ETLs analyzed above, equations (25) and (26) contain an additional parameter $\gamma$ (stiffness of the diagonal bond). Here, the obtained resonance frequencies that demarcate pass- and stop-bands are

$$(27) \quad \begin{aligned} &\gamma \in [0, 0.5]: \quad \omega_r = \sqrt{8} \quad (k_x = \pm k_y = \pm \pi), \\ &\gamma > 0.5: \quad \omega_r = (2\gamma + 1)/\sqrt{\gamma} \quad (k_x = \pm k_y = \pm \arccos(-1/2\gamma)) \end{aligned}$$

A simple LPW can be detected in RTLs by referring to a tri-diagonal strip constructed in such a manner that particles in the interior diagonal are in anti-phase oscillation and connected (*i*) with each other by bonds of the stiffness $\gamma$, and (*ii*) with immobile nodes in two outer diagonals by bonds with $g=1$. Then the total dimensionless stiffness of the RTL partial system is $4(1+\gamma)$, and the LPW frequency is $\omega = 2\sqrt{1+\gamma}$. This resonant frequency can be explicitly obtained from (26). The corresponding contours partially consist of straight lines $k_y = \pm\pi - k_x$, while other parts (if they are real) are curvilinear.

Thus, in the directions coinciding with the diagonal $y = x$, there exists a LPW with the same anti-phase oscillation form as in the square-cell MSL. The relations for the energy flux obtained for the RCL (with $l=1$) is also preserved for the considered RTL, if we set resonance frequency $\omega = 2\sqrt{1+\gamma}$ instead of $\omega = 2$ (corresponding to the MSL) in (11) and (12). Then we obtain similar expressions for group velocity projections and for the energy flux oriented along two directions $\beta = \pm\pi/4$:

$$(28) \quad (c_g)_x = \frac{\sin k_x}{2\sqrt{1+\gamma}}, \quad (c_g)_y = \frac{\sin k_y}{2\sqrt{1+\gamma}} = \frac{\sin k_x}{2\sqrt{1+\gamma}} = (c_g)_x \quad (k_y = \pm\pi - k_x)$$

and

$$(29) \quad |c_g| = \frac{\sqrt{2}}{2\sqrt{1+\gamma}} \sin \frac{\pi}{1+|\tan\alpha|} \quad \left(\alpha = \arctan\frac{k_y}{k_x}\right), \quad \beta = \pm\frac{\pi}{4}.$$

In Fig. 18, the resonant contours for three different values of $\gamma$ frequencies, $\gamma = 0.44$, 1 and $2.0625$, are shown by solid lines (corresponding resonant frequencies are $\omega = 2.4$, $\sqrt{8}$ and 3.5).

The main difference between the resonant LPW in RTL and RCL ($\gamma = 0$) is that in the former case LPWs exist only along the diagonals with additional bonds ($\gamma \neq 0$). Two symmetric straight sides $k_y = \pm\pi + k_x$ of the square contour inherent to the MSL are as if decomposed in the RTL case into parts, whose forms essentially depend on $\gamma$. For example, in the case of relatively small $\gamma = 0.44$, these parts are represented by sloping arcs in the second and forth quadrants of $k_x, k_y$-plane. The contour corresponding to $\omega = 3.5$ $(\gamma = 2.0625)$ consists of the two above-mentioned straight lines $k_y = \pm\pi - k_x$ and two sloping arcs located within the same, first and third, quadrants.

An intriguing pattern has been detected in the case of $\omega = \sqrt{8}$ ($\gamma = 1$): the mentioned additional part of the contour consists of straight lines $k_x = \pm\pi$, $k_y = \pm\pi$. Thus, in this case the resonant equifrequency contour determines six preferable directions of wave propagation, two of them along diagonals with additional bonds and four along the coordinate axes. Following (29), we obtain maximal velocities of energy flux along the six mentioned directions:

$$(30) \quad |c_g|_{max} = \frac{1}{2} \left(\beta = \pm\frac{\pi}{4}\right), \quad |c_g|_{max} = \frac{1}{\sqrt{8}} \left(\beta = \pm 0 \text{ and } \beta = \pm\frac{\pi}{2}\right)$$

To establish the average orientation of the group velocities in the considered cases $\omega = 2.4$ and $\omega = 3.5$, we approximate the above-mentioned arcs by secants (see dotted straight lines). In the mentioned cases, the energy flux orientations are determined by normals to these secants. Here, as in case of $\gamma = 1$, six preferable orientations of energy flux shown by bold arrows are observed.

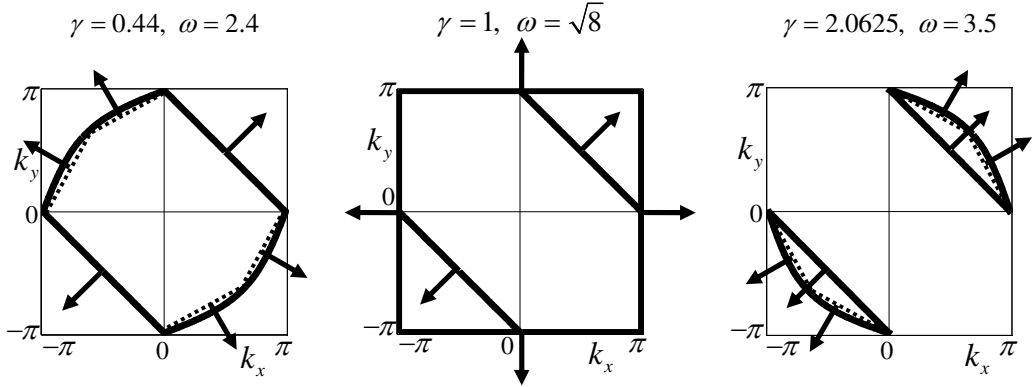

Fig. 18 Resonant contours. Arrows show energy flux orientations. Dotted straight lines are secants approximating curvilinear parts of contours.

## 5 CONCLUSIONS

Studies of mathematical models of wave propagation in 2D periodic infinite rectangular, triangular and hexagonal-cell lattices have been conducted. Eigenfrequencies of steady-state problems have been determined, and dispersion properties of free waves are described. We have shown that the frequency spectra of the models possess several resonant points located both at the boundary of pass/stop bands, and in the interior of a pass band. Special attention was given to the LPWs which exist at the resonance frequencies in the interior of pass-bands. As shown above, the existence of the LPW is a consequence of a certain symmetry of the lattice structure. It could be concluded that the existence of LPWs can be expected in more complicated 2D/3D lattices.

The boundary value problems with a local monochromatic source have been explored, and resonant patterns were revealed. The most important results include the analysis of particularities of transient waves in above-mentioned resonant cases, and the correspondence of the obtained wave beaming phenomena to the LPW analysis.

**ACKNOWLEDGMENTS**

This work was supported by The Israel Science Foundation, Grant 504/08